# About Algorithm for Transformation of Logic Functions (ATLF)

*Lev Cherbanski*

**Abstract.** In this article the algorithm for transformation of logic functions which are given by truth tables is considered. The suggested algorithm allows the transformation of many-valued logic functions with the required number of variables and can be looked in this sense as universal.

## 1 Introduction

There are not enough universal instruments for the work with logic functions. These functions are basis for computers, kryptography and other quickly developing areas of the technology. Historically the first effective mathematical apparatus for work with two-valued logic functions was the algebra developed by G. Boole. Boolean algebra is also the basis for diverse mathematical theories, as for example for Automated Theorem Proving. Łukasiewicz [1] and Post [2] extended the concept of binary logic and analysed different interpretations for many-valued logic functions. For these functions (3-valued, 4-valued etc.) Boolean algebra cannot be applied. Other attempts to develop an analogous systems were not successful.

## *2 Basic definitions*

Let's consider a discrete set B containing r elements. We name these elements with the symbols $b_1, b_2,...b_r$. An ordered set of n symbols comming from the elements of the set B is called a vector or a tuple. A n-tuple $(X_1,..,X_n)$ is an element of the discrete set $A_n$. The number of different n-tuples of the set $A_n$, which is formed on the basis of the set B, is equal $r^n$. The elements of the set $A_n$ can be ordered (or numbered) with different methods. Such an ordered set from n-tuples is the domain of logic functions from n variables. The range of logic functions is the set B. Here are the expressions which called logic functions:

$$f(X_1,..,X_n), \text{ where } X_1,...,X_n \text{ - are variables,} \qquad (1)$$

$$f(T_1,..,T_n), \text{ where } T_1,...,T_n \text{ - are logic functions.} \qquad (2)$$

## 3 Representation of logic functions

The first problem faced in working with logic functions is the representation of these functions. From the preceding considerations follows, that the order of the numbering of the n-tuples $(X_1,..,X_n)$ from the domain of logic functions is fixed or standardized. Then the accordingly ordered tuple of the values of a logic function is at the same time the shortest representation form of this function. Such tuple can be used without using the domain of function.

As an example we show in table 1 the logic function $y = f(X_1)$ from one 4-valued variable. To this purpose each of 1-tuples from the domain of function we put in conformity one value from the set B ~ {a,b,c,d}. One calls such a value table „a truth table".

| \ | $X_1$ | y |
|---|---|---|
| 0 | a | a |
| 1 | b | a |
| 2 | c | c |
| 3 | d | b |

Tab1. Representation of a 4-valued logic function $y = f(X_1)$ with n = 1.

The whole number of different logic functions with given n is limited and equal $r^{r^n}$. In our example the whole number of the functions with n = 1 is equal $4^{4^1}$, what $4^4$ is =256. We can present the function from the example in tab.1 also as a vector line (tuple):

   y= f($X_1$) ~ [b c a a].

With n> 5, or n> 6 the number of these symbols in the tuple is bigger than $r^5$, $r^6$ and in such cases it is necessary to store the data on a memory device.

**4 The calculation of the values of logic functions**

The second problem arising in working with logic functions is to define the value of the function presented in the form (1) corresponding some argument-tuple. In the case of a continuous function one calls this procedure „calculation of the values of a function with given value of the argument". Let's consider a case when the logic function is stored on a memory device. The algorithm for calculation of the values of the logic function consist of two steps:

- With the first step a numerical address is formed on the basis of the given argument-tuple for the storage space in the memory.
- In the second step the value of the function which is at this address is selected from the memory.

The algorithm of formation of the numerical address is defined by a rule of numbering of arguments-tuples, and for $r^n$ unique arguments-tuples should form $r^n$ unique addresses.

**5 Algorithm for transformation of logic functions (ATLF)**

The third and most important problem consists in transformation of logic functions, to be exact - a finding of functions from functions by definition (2). This procedure is called a composition of functions. Let's consider, how transformation of a kind is carried out.

We define    $y(X_1,..,X_n) = g(f_k(X_1,..,X_n))$  $k \in 1, 2, ... m,$    (3)

   where $g(Z_1,..,Z_m)$ - is the transforming logic function of m variables,

      $y(X_1,..,X_n)$ - is the resultant logic function from n variables,

      $f_k(X_1,..,X_n)$ $k \in 1, 2, ... m$ - is the ordered set from m logic function-arguments.

- For each argument-tuple ($X_1,..,X_n$) values of function-arguments $f_1,..,f_m$ are defined.
- The won m-tuple ($Z_1,..,Z_m$) is used as an argument-tuple of m-variables transforming logic function $g(Z_1,..,Z_m)$.
- The value of transforming function is calculated by means of the two-step-by-step procedure explained above in section 4. This value is the value required resultant function.
- Such calculations are spent for all $r^n$ tuples ($X_1,..,X_n$). The ordered tuple from $r^n$ values is the required resultant function $y(X_1,..,X_n)$.

# 6 Application example for the functional way of the algorithm

On the set {a, b, c} three 3-valued functions $f_1$, $f_2$, $f_3$ are given.

| \ | $X_1$ | $X_2$ | $f_1$ | $f_2$ | $f_3$ |
|---|---|---|---|---|---|
| 0 | a | a | a | a | c |
| 1 | a | b | c | c | c |
| 2 | a | c | c | b | b |
| 3 | b | a | b | a | c |
| 4 | b | b | b | c | c |
| 5 | b | c | a | a | c |
| 6 | c | a | a | b | b |
| 7 | c | b | a | a | c |
| 8 | c | c | b | c | b |

Tab.2 Tabular representation of logic function-arguments $f_1, f_2, f_3$ from two variables.

The 3-valued transforming function $g(Z_1, Z_2, Z_3)$ from three variables on the basis of the set {a, b, c} is given (Tab.3).

| \ | $Z_1$ | $Z_2$ | $Z_3$ | g | \ | $Z_1$ | $Z_2$ | $Z_3$ | g | \ | $Z_1$ | $Z_2$ | $Z_3$ | g |
|---|---|---|---|---|---|---|---|---|---|---|---|---|---|---|
| 0 | a | a | a | a | 9 | b | a | a | c | 18 | c | a | a | b |
| 1 | a | a | b | a | 10 | b | a | b | b | 19 | c | a | b | b |
| 2 | a | a | c | c | 11 | b | a | c | a | 20 | c | a | c | b |
| 3 | a | b | a | c | 12 | b | b | a | c | 21 | c | b | a | a |
| 4 | a | b | b | c | 13 | b | b | b | b | 22 | c | b | b | b |
| 5 | a | b | c | a | 14 | b | b | c | a | 23 | c | b | c | a |
| 6 | a | c | a | b | 15 | b | c | a | c | 24 | c | c | a | c |
| 7 | a | c | b | a | 16 | b | c | b | b | 25 | c | c | b | c |
| 8 | a | c | c | a | 17 | b | c | c | a | 26 | c | c | c | a |

Tab.3 Tabular representation of the function $g(Z_1, Z_2, Z_3)$. In the first column there is the numerical address of the value of the function **g** which was calculated from the suitable tuple.

It is necessary to find a composition of functions under the formula:

$$y(X_1, X_2) = g(\ f_1(X_1, X_2),\ f_2(X_1, X_2),\ f_3(X_1, X_2)\ ) \qquad (4)$$

For 2-tuple $(X_1,X_2) = [a,a]$ from the table 2 we calculate from functions $f_1, f_2, f_3$ 3-tuple [a a c]. In the table 3 the value of the function $g(a, a, c) = c$ corresponds to this tuple what gives y (a, a) = c. In the same way each of the nine 3-tuples, made from values of logic functions $f_2$ $f_1, f_3$, allows with the table of function $g(Z_1, Z_2, Z_3)$ to define the corresponding value of function $y(X_1, X_2)$ (tab. 4).

| \ | $X_1$ | $X_2$ | y |
|---|---|---|---|
| 0 | a | a | c |
| 1 | a | b | a |
| 2 | a | c | b |
| 3 | b | a | a |
| 4 | b | b | a |
| 5 | b | c | c |
| 6 | c | a | c |
| 7 | c | b | c |
| 8 | c | c | b |

Tab. 4  Tabular representation of the resultant function $y(X_1, X_2)$.

It is possible to present the function $y(X_1, X_2)$ also in the form of: [b c c c a a b a c].

The considered example also shows, that in expression (3) various numbers of variables n and m are possile: n ≥ m oder m ≥ n.

**7 Two variants of a reverse task setting**

Let's turn again to expression (3). In section 5 on the basis of this expression the direct task by definition of a composition of functions in the following statement is solved:

    The set of logic function-arguments  $f_k(X_1,..,X_n)$  is given.

    The transforming function  $g(Z_1,..,Z_m)$ is given.

    The values of the resultant function $y(X_1,..,X_n)$ are searched for.

On basis of expression (3) two reverse tasks can be formulate:

**7.1 The first variant of the reverse task**

    The set of logic function-arguments  $f_k(X_1,..,X_n)$  is given.

    The resultant function $y(X_1,..,X_n)$ is given.

    The values of the transforming function $g(Z_1,..,Z_m)$ are searched for.

The decision of the problem by such task setting is one or several functions.

The given sets of values of function-arguments and the corresponding values of the resultant function define completely a subset of values of a transforming function, which can be defined as *bounded*. All the other values of the transforming function can accept different values from the set B, and so are defined as not *bounded* ones.

One of the tuple from not bounded values together with all bounded values form an item of the transforming function which gives the solution of the problems for the reverse task.  The number of all possible combination of all not bounded values defines the whole number of the solutions of the problem. The subset of all not bounded values can be empty. In this case the problem has only one solution.

**7.2 The second variant of the reverse task**

The resultant function $y(X_1,..,X_n)$ is given.

The transforming function $g(Z_1,..,Z_m)$ is given.

One or several functions from the set of the logic function-arguments $f_k(X_1,..,X_n)$ is searched for.

This reverse task setting can come to several solutions (several different tuple of function-arguments). However, a solution can be absent for some combinations of resultant and transforming functions.

*Concluding remarks*

The suggested approach allows to solve the primary goals arising at work with logic functions of various valuedity, including with two-valued. Furthermore the offered algorithm of composition opens new areas for the application of logic functions on the base of the reverse tasks.

**References**

[1]   Łukasiewicz, Jan: Selected writings. North-Holland, 1970. Edited by L. Borowski.

[2]   Post, E.L.: Introduction to a general theory of elementary propositions. American Journal of Mathematics. Vol. 43 (1921), N. 3. P. 163-185. (republished: [Van Heijenoort J. (ed.) 1967]. P. 264-283).

[3]   Cherbanski, L.: Algorithm for Transformation of Logic Functions (ATLF). NG Verlag, Berlin, 2007. (In three languages.)

# Über den Algorithmus zur Transformation logischer Funktionen (ATLF)

*Lev Cherbanski*

In dem Artikel wird ein Algorithmus zur Umwandlung logischer Funktionen betrachtet, wobei die logischen Funktionen durch die gegebenen Wahrheitstabellen dargestellt werden. Der vorgeschlagene Algorithmus ermöglicht die Transformation mehrwertiger logischer Funktionen mit der benötigten Anzahl von Variablen und kann in diesem Sinne als universell betrachtet werden.

## 1   Einführung

In der Mathematik fehlt es an universellen Instrumenten für die Arbeit mit logischen Funktionen. Diese Funktionen dienen als Grundlage für Rechentechnik, Kryptographie und andere sich schnell entwickelnde Gebiete der Informationstechnik. Historisch gesehen war die erste mathematische Entwicklung für die Arbeit mit binären logischen Funktionen die Algebra, die von G. Boole entwickelt wurde. Boolsche Algebra ist ebenfalls die Grundlage für verschiedene mathematische Theorien, wie z.B. für automatische Theorembeweise. Łukasiewicz [1] und Post [2] erweiterten die Konzeption binärer Logik und betrachteten verschiedene Interpretationen für mehrwertige logische Funktionen. Für diese Funktionen (dreiwertige, vierwertige usw.) kann die Boolsche Algebra nicht angewendet werden. Weitere Versuche, ein analoges System zu entwickeln, führten nicht zum Erfolg.

## 2 Grunddefinitionen

Wir betrachten eine diskrete Menge B, welche r Elemente enthält. Wir bezeichnen diese Elemente mit den Symbolen $b_1, b_2,..,b_r$. Eine geordnete Menge von n Symbolen welche aus den Elementen der Menge B besteht, nennt man Vektor oder ein Tupel. Ein n-Tupel $(X_1,..,X_n)$ ist ein Element der diskreten Menge $A_n$. Die Anzahl verschiedener n-Tupel der Menge $A_n$, welche auf der Grundlage der Menge B gebildet wird, ist gleich $r^n$. Die Elemente der Menge $A_n$ können mit verschiedene Methoden geordnet bzw. durchnummeriert werden. Eine solche geordnete Menge von n-Tupel ist der Definitionsbereich logischer Funktionen von n Variablen. Der Wertebereich logischer Funktionen ist die Menge B. Als logische Funktionen werden die Ausdrücke bezeichnet, die wie folgt aussehen:

| | | |
|---|---|---|
| $f(X_1,..,X_n)$, wobei $X_1,..,X_n$ | – Variablen sind, | (1) |
| $f(T_1,..,T_n)$, wobei $T_1,..,T_n$ | – logische Funktionen sind. | (2) |

## 3  Darstellung logischer Funktionen

Die erste Aufgabe, die bei der Arbeit mit logischen Funktionen entsteht, ist die Darstellung dieser Funktionen. Wie aus den vorangegangenen Betrachtungen folgt, ist die Ordnung der Numeration der n-Tupel $(X_1,..,X_n)$ eines Definitionsbereiches festgelegt, bzw. standardisiert. Dann ist das entsprechend geordnete Tupel der Werte einer logischen Funktion gleichzeitig die kürzeste Darstellungsform dieser Funktion. Ein solches kann genutzt werden ohne den Definitionsbereich hinzuzuziehen. Als Beispiel stellen wir in der Tabelle 1 die logische Funktion $y= f(X_1)$ einer vierwertigen Variable dar. Dazu wird jeden 1-Tupel aus dem Definitionsbereich der Funktion ein entsprechender Wert aus der Menge B ~ {a,b,c,d} gegenübergestellt. Eine solche Wertetabelle nennt man Wahrheitstabelle.

| \ | $X_1$ | y |
|---|---|---|
| 0 | a | a |
| 1 | b | a |
| 2 | c | c |
| 3 | d | b |

Tab1. Tabellarische Darstellung einer vierwertigen logischen Funktion y=f($X_1$) bei n = 1.

Die gesamte Anzahl verschiedene logische Funktionen bei gegebenen n ist endlich und gleicht $r^{r^n}$. In unserem Beispiel bei n= 1 die gesamte Anzahl der Funktionen gleich $4^{4^1}$, was gleich $4^4$ = 256 ist. Die Funktion aus dem Beispiel in Tab.1 können wir ebenfalls als eine Vektorzeile (Tupel) darstellen:

$y = f(X_1) = [b\ c\ a\ a]$.

Bei n > 5, oder n > 6 ist die Anzahl dieser Symbole in solchem Funktionstupel größer als $r^5$, $r^6$ und bei solchen Fällen ist es notwendig die Daten auf einem Speichermedium zu lagern.

### 4 Die Bestimmung der Werte logischer Funktionen

Eine zweite Aufgabe, die sich bei der Arbeit mit logischen Funktionen stellt, ist Bestimmung der Werte der Funktion, die einem Argumententupel entspricht. Im Fall einer stetigen Funktion nennt man dieses Verfahren: Berechnung der Werte einer Funktion bei gegebenem Wert des Arguments. Betrachten wir einen Fall wenn eine logische Funktion auf einem Speichermedium gelagert wird. Der Algorithmus zur Bestimmung der Werte logischer Funktion besteht aus zwei Schritten:

1. Beim ersten Schritt wird auf der Grundlage der gegebenen Argumententupel eine numerische Adresse für den Speicherplatz auf dem Speichermedium gebildet.
2. Im zweitem Schritt wird aus dem Speichermedium der Wert der Funktion, welche sich unter dieser Adresse befindet ausgelesen.

Der Algorithmus zu Bildung von Adressen wird durch die Regel zur Nummerierung von Argumententupeln bestimmt und für $r^n$ von einzigartigen Argumententupeln müssen auch $r^n$ von einzigartigen Adressen gebildet werden.

### 5 Algorithmus zur Transformation von logischen Funktionen (ATLF)

Die dritte und die wichtigste Aufgabe besteht in der Transformation logischer Funktionen, oder exakter, zum Finden von der Funktion von Funktions-Argumenten nach der Definition (2). Dieses Verfahren nennt man Komposition von Funktionen.

Betrachten wir wie folgende Transformation ausgeführt wird:

$y(X_1,..,X_n) = g(f_k(X_1,..,X_n))\ k \in 1,2,...,m,$ (3)

wobei $g(Z_1,..,Z_m)$ - die transformierende logische Funktion von m Variablen ist;

$y(X_1,..,X_n)$ - die resultierende logische Funktion von n Variablen ist;

$f_k(X_1,..,X_n)\ k \in 1,2,..,m$ - der geordnete Satz aus m logischen Funktions-Argumenten ist.

- Für jedes Argumententupel $(X_1,..,X_n)$ werden die Werte der Funktions-Argumenten $f_1,..,f_m$ bestimmt.
- Der gewonnene m-Tupel $(Z_1,..,Z_m)$ wird benutzt als Tupel-Argument einer m-stelligen transformierenden logischen Funktion $g(Z_1,..,Z_m)$.
- Der Wert dieser Funktion wird bestimmt mit Hilfe des zweischrittiges Verfahrens das oben bei Punkt 4 erklärt wurde. Dieser Wert ist der Wert der gesuchten resultierenden Funktion.
- Solche Berechnungen führt man für alle $r^n$ Tupeln $(X_1,..,X_n)$ durch. Der geordnete Tupel aus $r^n$ Werten ist die gesuchte resultierende Funktion $y(X_1,..,X_n)$.

**6 Anwedungsbeispiel für die Funktionsweise des Algorithmus.**

Gegeben sei drei dreiwertige Funktionen $f_1, f_2, f_3$ auf der Menge {a, b, c} (Tab.2).

| \ | $X_1$ | $X_2$ | $f_1$ | $f_2$ | $f_3$ |
|---|---|---|---|---|---|
| 0 | a | a | a | a | c |
| 1 | a | b | c | c | c |
| 2 | a | c | c | b | b |
| 3 | b | a | b | a | c |
| 4 | b | b | b | c | c |
| 5 | b | c | a | a | c |
| 6 | c | a | a | b | b |
| 7 | c | b | a | a | c |
| 8 | c | c | b | c | b |

Tab.2 Tabellarische Darstellung logischer Funktions-Argumenten $f_1, f_2, f_3$ von zwei Variablen.

Die dreiwertige transformierende Funktion $g(Z_1,Z_2,Z_3)$ von drei Variablen auf der Grundlage der Menge {a, b, c} gegeben ist (Tab.3).

| \ | $Z_1$ | $Z_2$ | $Z_3$ | g | \ | $Z_1$ | $Z_2$ | $Z_3$ | g | \ | $Z_1$ | $Z_2$ | $Z_3$ | g |
|---|---|---|---|---|---|---|---|---|---|---|---|---|---|---|
| 0 | a | a | a | a | 9 | b | a | a | c | 18 | c | a | a | b |
| 1 | a | a | b | a | 10 | b | a | b | b | 19 | c | a | b | b |
| 2 | a | a | c | c | 11 | b | a | c | a | 20 | c | a | c | b |
| 3 | a | b | a | c | 12 | b | b | a | c | 21 | c | b | a | a |
| 4 | a | b | b | c | 13 | b | b | b | b | 22 | c | b | b | b |
| 5 | a | b | c | a | 14 | b | b | c | a | 23 | c | b | c | a |
| 6 | a | c | a | b | 15 | b | c | a | c | 24 | c | c | a | c |
| 7 | a | c | b | a | 16 | b | c | b | b | 25 | c | c | b | c |
| 8 | a | c | c | a | 17 | b | c | c | a | 26 | c | c | c | a |

Tab.3 Tabellarische Darstellung der Funktion g($Z_1$,$Z_2$,$Z_3$). In der ersten Spalte befindet sich die numerische Adresse des Wertes der Funktion, welche aus dem entsprechenden Variablentupel berechnet wurde.

Es sollte die Komposition von Funktionen nach der Formel

$$y(X_1,X_2) = g(f_1(X_1,X_2), f_2(X_1,X_2), f_3(X_1,X_2)). \qquad (4)$$

bestimmt werden.

Für den 2-Tupel ($X_1$,$X_2$) = [a  a] aus der Tabelle 2 bekommen wir für Funktionen $f_1$, $f_2$, $f_3$ ein 3-Tupel der Werte [a a c]. In der Tabelle 3 entspricht ihm der Wert der Funktion g(a, a, c) =c, was ergibt y(a, a) = c. Nach dem selben Verfahren kann aus jedem der neun 3-Tupel, die aus den Werten der logischen Funktionen $f_1$, $f_2$, $f_3$, bestehen, mit Hilfe der Funktionstabelle von g($Z_1$,$Z_2$,$Z_3$) der gesuchte Wert der Funktion y($X_1$,$X_2$) bestimmt werden (Tab. 4).

| \ | $X_1$ | $X_2$ | y |
|---|---|---|---|
| 0 | a | a | c |
| 1 | a | b | a |
| 2 | a | c | b |
| 3 | b | a | a |
| 4 | b | b | a |
| 5 | b | c | c |
| 6 | c | a | c |
| 7 | c | b | c |
| 8 | c | c | b |

Tab. 4  Tabellarische Darstellung der Funktion y($X_1$,$X_2$).

Die Funktion y($X_1$,$X_2$) kann man auch so darstellen: [b c c c a a b a c].

Das dargestellte Beispiel zeigt auch, dass  im Ausdruck (3) Funktionen mit unterschiedlicher Anzahl von Variablen n und m  möglich sind:  n ≥ m oder m ≥ n.

### 7 Zwei Varianten der Aufgabenstellung zur umgekehrten Aufgabe

Wenden wir uns wieder dem Ausdruck (3) zu. Im 5. Abschnitt wurde auf der Grundlage dieses Ausdrucks die Aufgabe zu Bestimmung der Komposition von Funktionen in folgende Aufgabenstellung gelöst:

Gegeben ist die Menge logischer Funktions-Argumente $f_k(X_1,..,X_n)$.

Gegeben ist die transformierende Funktion g($Z_1$,..,$Z_m$).

Gesucht sind die Werte der resultierende Funktion y($X_1$,..,$X_n$).

Auf Grundlage von Ausdruck (3) können zwei umgekehrte Aufgaben gestellt werden:

### 7.1 Erste Variante der umgekehrten Aufgabe

Gegeben ist der Tupel von logischen Funktions-Argumenten $f_k(X_1,..,X_n)$.

Gegeben ist die resultierende Funktion $y(X_1,..,X_n)$.

Gesucht sind die Werte der transformierenden Funktion $g(Z_1,..,Z_m)$.

Als Lösung der Aufgabe mit einer solche Aufgabenstellung gibt es eine oder mehrere Funktionen.

Die vorhandenen Werttupel von Funktions-Argumenten und die Werte der resultierenden Funktion werden vollständig die Untermenge der Werte der transformierende Funktion bestimmen, die man als gebunden definieren kann. Alle anderen Werte der transformierenden Funktion können unterschiedliche Werte aus der Menge B annehmen, und sind deswegen als nicht gebundene definiert.

Die Zusammenstellung aus den gewonnenen gebundenen Werten zusammen mit den freigewählten nicht gebundenen Werten ergeben ein Exemplar der transformierenden Funktion, was die Lösung der umgekehrten Aufgabe ist. Werte die, der resultierenden Funktion entsprechen, welche verschiedene Werte aus der Menge B annehmen könnte. Die Anzahl der möglichen Kombinationen aller nicht gebundenen Werte definiert die gesamte Anzahl der Aufgabenlösungen. Die Untermenge aller nicht gebundener Werte kann leer sein. In diesem Fall hat die Aufgabe nur eine Lösung.

### 7.2 Zweite Variante der umgekehrten Aufgabe

Gegeben ist die resultierende Funktion $y(X_1,..,X_n)$.

Gegeben ist die transformierende Funktion $g(Z_1,..,Z_m)$.

Gesucht ist eine oder mehrere Funktionen aus dem Tupel der logischen Funktions-Argumenten $f_k(X_1,..,X_n)$.

Diese umgekehrte Aufgabestellung kann mehrere Lösungen besitzen (Mehrere unterschiedliche Tupel von Funktions-Argumenten). Jedoch kann für einige Kombinationen von resultierenden und transformierenden Funktionen eine Lösung komplett fehlen.

### Schluss

Die vorgeschlagene Vorgehensweise erlaubt es grundlegende Aufgaben zu lösen, welche bei der Arbeit mit logischen Funktionen von unterschiedlichen Wertigkeiten, auch zweiwertigen, entstehen können. Weiterhin kann der vorgeschlagene Algorithmus für Komposition (ATLF) neue Gebiete für die Anwendung logischer Funktionen auf der Basis der umgekehrten Aufgaben öffnen.

### Literatur

[1] J. Łukasiewicz. Jan Łukasiewicz, *Selected writings*. North-Holland, 1970. Edited by L. Borowski.

[2] Post E.L. 1921, *Introduction to a general theory of elementary propositions* //

American Journal of Mathematics. Vol. 43, N. 3. P. 163-185. (Republisht: [Van Heijenoort J. (ed.) 1967]. P. 264-283).

[3] Cherbanski, L.: Algorithm for Transformation of Logic Functions (ATLF). NG Verlag, Berlin, 2007. (In three languages.)

# Об алгоритме преобразования логических функций

*Лев Щербанский*

В статье рассматривается алгоритм для преобразования логических функций, заданных таблицами истинности. Предложенный алгоритм обеспечивает преобразование логических функций различной значности с необходимым количеством переменных и в этом смысле является универсальной.

## 1  Введение

В математике отсутствуют универсальные инструменты для работы с логическими функциями. Логические функции служат основой для вычислительной техники, криптографии и других быстроразвивающихся областей техники. Исторически первым эффективным математическим аппаратом для работы с двузначными логическими функциями явилась алгебра, разработанная Дж. Булем. Булева алгебра является также основой различных математических теорий, таких как автоматическое доказательство теорем. Лукашевич [1] и Пост [2] расширили концепцию двузначной логики и рассмотрели различные интерпретации многозначных логических функций. Для многозначных логических функций (троичных, четверичных и пр.) булева алгебра неприменима и попытки разработать аналогичный аппарат к успеху не привели.

## 2  Основные определения

Рассмотрим дискретное множество B, содержащее r элементов. Обозначим эти элементы символами $b_1, b_2,...,b_r$. Упорядоченную совокупность символов, составленную из элементов множества B, называют *вектором* или *набором*. Набор из n символов $(X_1,..,X_n)$ является элементом дискретного множества $A_n$. Число различных наборов множества $A_n$, формируемых на базе множества B, равно $r^n$. Элементы множества $A_n$ можно упорядочить (пронумеровать) различными методами. Такое упорядоченное множество наборов является *областью определения логических функций от n переменных*. Областью значений логических функций является множество B. *Логическими функциями* называются выражения одного из следующих видов:

$f(X_1,...,X_n)$, где $X_1,..,X_n$ – переменные, (1)

$f(T_1,...,T_n)$, где $T_1,..,T_n$ – логические функции. (2)

## 3  Представление логических функций

Первая задача, возникающая при работе с логическими функциями - это представление этих функций. Как следует из предыдущих рассуждений, если порядок нумерации наборов $(X_1,..,X_n)$ области определения закреплен (стандартизован), то соответствующий упорядоченный набор значений логической функции является одновременно самым экономным представлением этой функции. Такой набор может использоваться без привлечения области определения.

В качестве примера представим таблицей 1 логическую функцию $y= f(X_1)$ от одной четырехзначной переменной. Для этого каждому из $4^1$ наборов из области определения функции поставим в соответствие одно значение из множества $B \sim \{a,b,c,d\}$. Такая таблица значений логической функции называется таблицей истинности.

| \ | $X_1$ | y |
|---|---|---|
| 0 | a | a |
| 1 | b | a |
| 2 | c | c |
| 3 | d | b |

Таб. 1. Табличное представление одной из четырехзначных логических функций $y=f(X_1)$ при n = 1.

Общее число различных логических функций при заданном n конечно и равно $r^{r^n}$. В нашем примере при n = 1 общее число функций равно $4^{4^1} = 4^4 = 256$. Функцию из примера Таб.1 можно также представить в виде вектора-строки:

$y = f(X_1) =$ [b c a a].

При n >5, n>6 число символов в таких наборах-функциях превышает $r^5$, $r^6$, и эти наборы-функции необходимо хранить в запоминающих устройствах.

### 4 Вычисление значений логических функций

Вторая задача, возникающая при работе с логическими функциями, - определение значения функции (1), соответствующего некоторому набору-аргументу. В случае непрерывных функций эта процедура называется вычислением значения функции при заданном значении аргумента. Рассмотрим случай, когда логическая функция хранится в запоминающем устройстве. Алгоритм определения значения логической функции состоит из двух шагов.

(1)  На первом шаге на основе заданного набора-аргумента формируется число-адрес участка памяти запоминающего устройства.

(2)  На втором шаге из памяти извлекается значение функции, хранящееся по этому адресу.

Алгоритм формирования адреса определяется правилом нумерации наборов-аргументов, и для $r^n$ уникальных наборов-аргументов должен формировать $r^n$ уникальных адресов.

### 5 Алгоритм нахождения функций от логических функций

Третья и наиболее важная задача состоит в преобразовании логических функций, а точнее - нахождении функций от функций по определению (2). Эту процедуру называют композицией функций.

Рассмотрим, как выполняется преобразование вида:

$y(X_1,..,X_n) = g(f_k(X_1,..,X_n))$ $k \in 1,2, ...m$ , (3)

где  $g(Z_1,..,Z_m)$ - преобразующая логическая функция от m переменных,

$y(X_1,..,X_n)$ – результирующая логическая функция от n переменных,

$f_k(X_1,..,X_n)$ $k \in 1,2,...m$ - упорядоченная совокупность из m логических

функций-аргументов от n переменных.

- Для каждого набора-аргумента $(X_1,..,X_n)$ определяются значения функций-аргументов $f_1,..,f_m$.
- Полученный набор $(Z_1,..,Z_m)$ используется как набор-аргумент m-местной преобразующей логической функции $g(Z_1,..,Z_m)$.
- С помощью двухшаговой процедуры, определенной выше в разделе 4, вычисляется значение этой функции.
- Такие вычисления проводятся для всех $r^n$ наборов $(X_1,..,X_n)$. Упорядоченный набор из $r^n$ значений является искомой результирующей функцией $y(X_1,..,X_n)$.

## 6    Пример применения алгоритма

Заданы на множестве {a, b, c} три трёхзначные функции $f_1$, $f_2$, $f_3$ ( Таб.2).

| \ | $X_1$ | $X_2$ | $f_1$ | $f_2$ | $f_3$ |
|---|---|---|---|---|---|
| 0 | a | a | a | a | c |
| 1 | a | b | c | c | c |
| 2 | a | c | c | b | b |
| 3 | b | a | b | a | c |
| 4 | b | b | b | c | c |
| 5 | b | c | a | a | c |
| 6 | c | a | a | b | b |
| 7 | c | b | a | a | c |
| 8 | c | c | b | c | b |

Таб. 2 Табличное представление логических функций-аргументов $f_1$, $f_2$, $f_3$ от двух переменных.

На множестве {a, b, c} задана также трёхзначная преобразующая функция $g(Z_1,Z_2,Z_3)$ от трех переменных. ( Таб.3).

| \ | $Z_1$ | $Z_2$ | $Z_3$ | g | \ | $Z_1$ | $Z_2$ | $Z_3$ | g | \ | $Z_1$ | $Z_2$ | $Z_3$ | g |
|---|---|---|---|---|---|---|---|---|---|---|---|---|---|---|
| 0 | a | a | a | a | 9 | b | a | a | c | 18 | c | a | a | b |
| 1 | a | a | b | a | 10 | b | a | b | b | 19 | c | a | b | b |
| 2 | a | a | c | c | 11 | b | a | c | a | 20 | c | a | c | b |
| 3 | a | b | a | c | 12 | b | b | a | c | 21 | c | b | a | a |
| 4 | a | b | b | c | 13 | b | b | b | b | 22 | c | b | b | b |
| 5 | a | b | c | a | 14 | b | b | c | a | 23 | c | b | c | a |
| 6 | a | c | a | b | 15 | b | c | a | c | 24 | c | c | a | c |
| 7 | a | c | b | a | 16 | b | c | b | b | 25 | c | c | b | c |
| 8 | a | c | c | a | 17 | b | c | c | a | 26 | c | c | c | a |

Таб. 3 Табличное представление функции $g(Z_1,Z_2,Z_3)$. В первом столбце расположен номер-адрес значения функции, вычисленный по соответствующему набору.

Необходимо найти композицию функций по формуле:

$$y(X_1,X_2) = g(f_1(X_1,X_2), f_2(X_1,X_2), f_3(X_1,X_2)). \qquad (4)$$

Для набора-аргумента $(X_1,X_2) = [a\ a]$ из таблицы 2 для функций $f_1, f_2, f_3$ получаем набор значений [a a c]. В таблице 3 ему соответствует значение функции $g(a,a,c) = c$, что даёт $y(a,a) = c$. Таким же образом каждый из девяти наборов, составленный из значений логических функций $f_1, f_2, f_3$, позволяет по таблице функции $g(Z_1,Z_2,Z_3)$ определить соответствующее значение функции $y(X_1,X_2)$.

| \ | $X_1$ | $X_2$ | y |
|---|---|---|---|
| 0 | a | a | c |
| 1 | a | b | a |
| 2 | a | c | b |
| 3 | b | a | a |
| 4 | b | b | a |
| 5 | b | c | c |
| 6 | c | a | c |
| 7 | c | b | c |
| 8 | c | c | b |

Таб. 4 Табличное представление функции $y(X_1,X_2)$.

Функцию $y(X_1,X_2)$ можно также представить в виде упорядоченного набора: [b c c c a a b a c]. Рассмотренный пример показывает, что в выражении (3) возможны различные соотношения между размерностями областей определения функций y и g: $n \geq m$ или $m \geq n$.

### 7  Два варианта постановки обратной задачи

Обратимся снова к выражению (3). В разделе 5 на основе этого выражения решается задача по определению композиции функций в следующей постановке:

Задана совокупность логических функций-аргументов $f_k(X_1,..,X_n)$.

Задана преобразующая функция $g(Z_1,..,Z_m)$.

Необходимо определить значения результирующей функции $y(X_1,..,X_n)$.

На основе этого выражения можно сформулировать также две обратные задачи.

### 7.1 Первый вариант обратной задачи

Задан набор функций-аргументов $f_k(X_1,..,X_n)$.

Задана результирующая функция $y(X_1,..,X_n)$.

Необходимо найти значения преобразующей функции $g(Z_1,..,Z_m)$.

Решением задачи в такой постановке будет одна или несколько функций. Заданные наборы значений функций-аргументов и соответствующие им значения результирующей функции полностью определяют подмножество значений преобразующей функции, которые можно определить как связанные. Остальные значения преобразующей функции могут принимать различные значения из множества B и поэтому называются свободными. Один из наборов свободных значений совместно со связанными значениями даёт экземпляр преобразующей функции, являющийся решением обратной задачи.

Число возможных комбинаций всех свободных значений определяет общее число решений задачи. Подмножество свободных значений может быть пустым. В этом случае задача имеет только одно решение.

### 7.2 Второй вариант обратной задачи

Задана результирующая функция $y(X_1,..,X_n)$.

Задана преобразующая функция $g(Z_1,..,Z_m)$.

Необходимо определить одну или несколько функций из совокупности $f_k(X_1,..,X_n)$.

Обратная задача в такой постановке может иметь несколько решений (несколько различных комплектов функций-аргументов). Однако для некоторых сочетаний результирующей и преобразующей функций решение может отсутствовать.

### Заключение

Предложенный подход позволяет решать основные задачи, возникающие при работе с логическими функциями различной значности, в том числе и с двузначными. Кроме того предложенный алгоритм нахождения композиции открывает новые области применения логических функций за счёт решения обратных задач.

### Литература

[1] J. Łukasiewicz. Jan Łukasiewicz, *Selected writings*. North-Holland, 1970. Edited by L. Borowski.

[2] Post E.L. 1921, *Introduction to a general theory of elementary propositions* //

American Journal of Mathematics. Vol. 43, N. 3. P. 163-185. (Republisht: [Van Heijenoort J. (ed.) 1967]. P. 264-283).

[3] Cherbanski, L.: Algorithm for Transformation of Logic Functions (ATLF). NG Verlag, Berlin, 2007. (In three languages.)